\documentclass[10pt, a4paper]{article}
\usepackage{lrec}
\usepackage{multibib}
\newcites{languageresource}{Language Resources}
\usepackage{graphicx}
\usepackage{tabularx}
\usepackage{soul}

\usepackage{epstopdf}
\usepackage[utf8]{inputenc}
\usepackage[english]{babel} 

\usepackage{hyperref}
\usepackage{xstring}

\title{Alexa as a CALL platform for children: Where do we start?}

\name{Nikos Tsourakis$^1$, Manny Rayner$^1$, Hanieh Habibi$^1$\\
\textbf{\large Pierre-Emmanuel Gallais}$^2$, \textbf{\large Cathy Chua}$^2$, \textbf{\large Matt Butterweck}$^2$}

\address{(1) Geneva University \\
        (2) Independent researcher \\
         \{Nikolaos.Tsourakis,Emmanuel.Rayner,Hanieh.Habibi\}@unige.ch\\
         cathyc@pioneerbooks.com.au, matthias@butterweck.de, gallais2009@hotmail.fr\\}

\abstract{Amazon's Alexa is now widely available and shows interesting potential as a platform for hosting CALL games aimed at children. In this paper, we describe an initial informal experiment where we created some simple CALL games and made them available to a few child testers. We report the children's and parents' reactions. Our overall conclusion is that, although Alexa has many positive features, there are still fundamental platform issues in the current version that make it very difficult to build compelling CALL games for children. The games used will soon be freely available for download on the Alexa store.\\ \newline }

\begin{document}

\maketitleabstract

\section{Introduction}

In the four years since its release, Amazon's Alexa has become a major platform for developing and deploying spoken language applications; according to Amazon, over one hundred million Alexa-enabled devices have now been sold. Amazon's advertising highlights the attractiveness of the platform to children, and one only needs to spend ten minutes watching a couple of kids playing with Alexa to see that this is not all hype. The device clearly has good acoustic models for children's speech; the far-field recognition and hands-free operation work well, allowing children to do other things while talking to the device; and the default ``always-on'' mode eliminates start-up time. Further attractive properties include a powerful and well-maintained API for developing and fielding applications (``skills''), and excellent scalability. We have for some time been developing a speech-enabled CALL platform \cite{RaynerEASlate2015Platform} which among other things has been used to build CALL games for children \cite{BaurEA2013ICERI,Baur2015}, and were curious to find out what we could do if we ported some of this functionality to Alexa.

Other differences when compared with conventional platforms also seemed potentially positive. The smartphone revolution in particular has paved the way for an interaction paradigm which proves to be a minefield of distractions, dominated by social media applications whose primary goal is to occupy as much of the user’s time as possible. For example, according to a dscout Mobile Touches study,\footnote{\url{https://blog.dscout.com/mobile-touches}} smartphone users on average touch, swipe or tap their phone over 2,500 times a day. The situation is no better on a desktop PC, where similar distractions are available. The interaction paradigm inherent in the Echo, in contrast, emphasises quick and purposeful interactions; this makes it an attractive candidate platform for child-oriented CALL applications, given that children are notoriously prone to distraction. Turn taking between the device and the user is normally restricted to the context of the ‘skill’, without being affected by other platform events. Finally, it offers a communal experience where multiple members of the same family or friends can interact with the device. Children do not look at an individual screen and, other things being equal, will find it easier to collaborate with others than they would if they were using a smartphone or tablet. 

Here, we report an initial experiment where we created a few CALL games and gave them to some children we were in contact with to see what happened. We recruited six children --- coincidentally, all boys --- aged between four and ten years old and belonging to four separate families, and gave an Echo Dot device to each family. A set of instructions was provided describing how to use the applications. Our unashamedly anecdotal analysis (you have to start somewhere) is based on observations while the subjects interacted with the system and from informal discussions with children and their parents. For reasons of space, we will focus on the three most active users, referred to here as HK, JK and VT.

The rest of the paper is structured as follows. In the next section, we describe the CALL games used. \S\ref{Section:Experiments} presents the results. The final section concludes.

\section{Alexa games used}
\label{Section:Games}


We began by creating five simple games. Each game was constructed in three versions, for English, German and French. The games, listed in Table~\ref{Table:AlexaGames}, will be available for free download from the Alexa store by May 15 2019. The structure of each game is the same; the basic strategy is prompt/response, where the prompt is either a recorded audio file (the games ``Which movie?'', ``Which language?'' and ``Which animal?''), or a piece of text spoken using Alexa's TTS functionality (the games ``Number game'' and ``Letter game''). Each game was first developed for English, then ported to German and French by native speakers.

\begin{figure}[!h]
\begin{center}
\includegraphics[width=8cm]{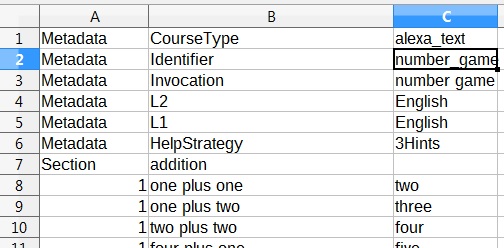}
\caption{Top of spreadsheet defining ``Number game''}
\label{figure:AnimalGame}
\end{center}
\end{figure}

In some of the games, prompts are divided into ``lessons''. In the arithmetic game, there are four lessons, for addition, subtraction, multiplication and division. In the animal noises and language ID games, there are two lessons called ``easy'' and ``difficult''. A lesson can optionally be further divided into numbered ``groups'', with the convention that prompts from the low-numbered groups are presented before prompts from the high-numbered groups. 
Games are defined using a simple spreadsheet format. The first few lines of the spreadsheet contain metadata (invocation phrase, L1, L2, etc); the body consists of lines defining prompt/response pairs, where the first column gives the group number, the second the prompt, and the third the permitted responses. Figure~\ref{figure:AnimalGame} gives an example of a game spreadsheet.

At each turn, the game speaks a prompt randomly chosen from the currently active group and lesson, with the proviso that prompts do not repeat in the same session. The player can respond immediately, in which case the game either confirms and moves on to the next turn if it judges the response correct, or else repeats the prompt if it judges it incorrect. In both cases, the game echoes back the player's response, using different strategies for correct and incorrect responses. If the response is judged incorrect, the echoed content is ``I heard...'' followed by the speech hypothesis from the Alexa recogniser. If the response is judged correct, the echoed content is the matched phrase from the spreadsheet.
Instead of responding directly to the prompt, the player may also use one of the following navigation commands:

\begin{description}

\item[Help:] Give a choice of three possible answers the first time; say the correct answer the second time.

\item[Repeat:] Repeat the last thing the app said.

\item[Wait:] Give the player more time to think. 

\item[Next:] Skip to the next prompt.

\item[Back:] Return to the previous prompt.

\item[Next lesson:] Skip to the next lesson.

\item[Lesson X:] (or simply ``X''). Skip to the lesson called ``X'', e.g. ``Lesson: addition'' or ``Addition'' goes to the lesson labelled ``addition'' in the spreadsheet.

\item[Lessons:] List the names of the available lessons.

\end{description}

In initial versions of the games, the decision as to whether a response was correct or incorrect was made by performing a simple comparison between the recognition hypothesis and the set of correct answers defined by the spreadsheet. Feedback from the children suggested that many of them were frustrated by the apps' brittleness; in later versions, we used the robust matching method from \cite{RaynerEA2017}, which anecdotally gives much more user-friendly performance.

\begin{table*}[ht!]
\begin{center}
\caption{Main Alexa games used in experiment. All games are available from the Alexa Store and free to download; search for the name of the game in one of the first three columns.}
\label{Table:AlexaGames}
\begin{tabular}{|c|c|c|c|l|}
\hline
\multicolumn{3}{|c|}{Name of the game}                           &\#Prompts   &Short description\\
\hline
English            &French         &German           &            &                  \\
\hline
Which movie?       &Quel film?     &Welcher Film?    &76          &Guess the movie from a short clip\\
\hline
Which language?    &Quelle langue? &Welche Sprache?  &88          &Guess the language from a short phrase\\
\hline
Which animal?      &Quel animal?   &Welches Tier?    &25          &Guess the animal from its sound\\
\hline
Number game        &Jeu de chiffres&Zahlenspiel      &100         &Practice spoken arithmetic\\
\hline
Letter game        &Jeu de lettres &Buchstabenspiel  &40          &Name things starting with a given letter\\
\hline
\end{tabular} 
\end{center}
\end{table*}

\subsection{Personalised courses: ``Pokémon''}

The courses described above are all basically generic ones, though to some extent they were designed with the children's interests in mind. (One of the children, living in highly multilingual Geneva, is very interested in languages and did indeed like the language ID game). During the course of the experiment, it occurred to us that it would be possible to go a step further and design courses that were explicitly personalised to a single student. We explored two versions of this idea.

In the first experiment, we invited one of the subjects, a bilingual English/French seven year old boy we will call HK, to design his own game. HK is a passionate devotee of Pokémon, and this seemed like a natural subject. The protocol for constructing the game was devised by his mother, AK, who acted as coordinator and secretary.

In AK's scenario, HK and his friend DM, a francophone boy of about the same age, sat facing each other, with one boy holding a deck of Pokémon cards oriented so that only he could see them; the two boys alternated roles. At each turn, the boy with the cards picked a card and made up a French question based on the card's text. The other boy tried to guess the Pokémon. After some discussion, they agreed on a question which AK wrote down in the spreadsheet. At the end of the session, AK mailed the spreadsheet to us, and we compiled and deployed the game. 

A couple of days later, HK and DM tried out their game. The initial reaction was very positive (they were amazed that their content had been turned into this new form), but they rapidly lost interest. There were several problems: in addition to the generic usability issues discussed in the next section, the game was, unsurprisingly, not very well designed, with questions that were both overlong and often too difficult even for Pokémon experts. There was also not enough content --- HK and DM only managed to generate a dozen questions before getting bored --- and a couple of the Pokémon names hardly ever got recognised. AK encouraged the children to try and identify the problems themselves and redesign the game. They produced a second version, with somewhat shorter prompts, one of the hard-to-recognise questions removed, and a little more content; but enough of the problems remained that they soon lost interest again, and could not be persuaded to produce a third version. 

\subsection{Personalised courses: ``V's homework''}
\label{Section:VsHomework}

In the second personalised course the target child, VT, was not part of the development loop. The basic motivation stemmed from the actual need of the child to practise small dialogues at home as part of his homework. Specifically, in French-speaking Switzerland children start learning German at school at the age of eight. During the first few months of the course, they are asked to develop their generation and comprehension skills by participating in small dialogues where they alternate the two roles. Normally, one of the  parents plays the role of the conversation partner. It occurred to us that it would be easy to adapt the Alexa framework described above and create a course that included a set of these small dialogues. The basic interaction pattern for a turn is as follows:

\begin{enumerate}

\item \textbf{Party 1}: \textless poses a question in German\textgreater
\item \textbf{Party 1}: \textless gives a hint answer in French\textgreater
\item \textbf{Party 2}: \textless responds in German\textgreater

\end{enumerate}

We used a slightly modified version of the spreadsheet format described above to define the course, making each dialogue into a ``lesson''. Prompts were realised as before using Alexa's TTS voice, with the L1 ``hint'' part marked up to be spoken more quietly. An example dialogue is shown in Table~\ref{Table:Vassili}. 

\begin{table*}[ht!]
\begin{center}
\caption{Sample dialogue for ``V's homework''. In each turn, the first element is the German phrase spoken by the app, the second element is the French ``hint'' spoken by the app, and the third element is an example of a correct response.}
\label{Table:Vassili}
\begin{tabular}{|l|c|c|r|}
\hline
Party  &Interaction                                      &(English gloss)\\
\hline
1	&Guten Morgen                                        &(Good morning)\\
1	&Bonjour (le matin)                                  &(Good morning)\\
2	&Guten Morgen                                        &(Good morning)\\
\hline
1	&Wie heißt du?                                       &(What is your name?)\\
1	&Je m'appelle V                                      &(My name is V)\\
2	&Ich heiße V                                         &(My name is V)\\
\hline
1	&Was möchtest du kaufen?                             &(What would you like to buy?)\\
1	&J'aimerais du fromage et de la limonade, s'il vous plaît&(I would like some cheese and some lemonade please)\\	
2	&Ich möchte Käse und Limonade bitte                  &(I would like some cheese and some lemonade please)\\
\hline 
1	&Bitte                                               &(There you are)\\
1	&Merci                                               &(Thank you)\\
2	&Danke                                               &(Thank you)\\
\hline
\end{tabular} 
\end{center}
\end{table*}


\section{Experimenting with Alexa}
\label{Section:Experiments}


\subsection{Feedback from AK, HK and JK}

The most diligent users in the study were definitely AK and her two children, HK and JK (7 and 11 year old boys). The family also encouraged several of the children's friends to try out the Alexa games when visiting.

We interviewed AK, HK and JK to get their impressions, and watched the two boys using the games. It was immediately obvious that they had mastered the technical problems of interacting with the games, and had played them enough that they knew the content quite well. Unfortunately, our impression was that they had not in fact used the games as CALL tools, but only as entertainment. As already noted, the boys, members of an English family who have grown up in French-speaking Geneva, are bilingual English/French. They had almost exclusively used the English and French versions of the games and hardly tried the German ones at all, despite the fact that JK had done a year of German at school and might well have benefited from using the German versions.

We are not sure we know why HK and JK were reluctant to use the games for an educational purpose, but it certainly seemed possible that this is related to a current misfeature of Alexa: the device language can only be changed from the web control panel. It \textit{cannot} be changed through a voice command. Since accessing the control panel requires the Amazon password, which AK was unwilling to give to her children, they could not activate the German versions of the games without asking AK for assistance; they would then have to ask for help a second time at the end of the session to switch back to French, which was the default interaction language. In short, the kids had no autonomy. They complained explicitly about this.

\subsection{Feedback from VT}

In contrast to HK and JK, VT used his Alexa device mostly for educational purposes, the ``V's homework'' course from \S~\ref{Section:VsHomework}.
Having previously been exposed to the content, it was straightforward for VT to complete the task, although it was obvious that performing the interaction with one of the parents was more engaging. 
VT also tried out all the games for his L1 (again French). He seemed genuinely interested in experimenting with this new gadget and continued to play until specifically told to stop. After the session, however, he did not ask to play with the device again.

\subsection{Common feedback}

Three generic problems were apparent, and the subject of repeated complaints from all subjects. First, Alexa is currently unable to handle barge-in.
Since children tended to interrupt the spoken output of the device anyway, we introduced a distinctive sound that signifies the start of each turn. The children had no trouble understanding the purpose of the earcon and interaction worked much better once it was introduced, but they did not like being forced to wait until the game had finished speaking before they could respond, and said that the games were ``too slow''. Shortening the prompts as much as possible did not correct the problem. 

In the opposite direction, Alexa also drops out of the game and returns to top level if the user stops speaking for more than a few seconds. Here, the best fix we could come up with was to introduce a ``wait'' command (essentially an extra turn), which again improved the situation. Nonetheless, the bottom line was that when the children knew the answer, they were not allowed to give it at once, and when they didn't know it, they were not allowed to pause freely, but had to remember to say ``wait''. In addition, although Alexa's speech recognition is very good by current standards, it was not perceived as being good enough; misrecognitions added to the general feeling of frustration. 

Despite this, all parents, in particular AK, stressed that they saw a great deal of positive potential in Alexa, and hoped that later versions of games like the ones we gave them would be able to realise that potential. It seems, however, that the current platform has too many negative aspects for CALL games like ours to work well with children in an unsupervised home setting.

\subsection{Feedback from group session on Open Day}
\label{GroupFeedback}

We carried out a short but interesting experiment in early November, 2018 in connection with ``Futur en tous genres'', a yearly Open Day organised for children of University of Geneva employees. A group of a dozen children, aged between ten and twelve, were scheduled to visit our lab and spend an hour interacting with our CALL software.

We had only two Alexa devices available; given this limitation, the protocol we decided to try was the following. The Alexa device was placed on a table, with the children grouped around it in a semi-circle. The first child was handed a token; the group was told that only the child with the token was allowed to speak, and after speaking was obliged to hand the token to their right-hand neighbour. We then launched several of the French and English versions of the games, and let the kids interact with them.

Somewhat to our surprise, this setup was very successful. The children followed the instructions without complaining, and gave every evidence of having a good time. There was a lot of smiling and laughing, and when someone got stuck they often received good-natured whispered help. Alexa's recognition functioned well, and things progressed smoothly, with rapid passing of the token. Our impression was that our guests were disappointed when the hour was up and could happily have stayed longer. 

\subsection{Social aspects}

Finally, some general remarks. First, when children used the device with other people --- family, friends or fellow students --- they seemed far more engaged in the gameplay. Conversely, when they were asked to play the games alone they were less motivated. This is consistent with the view that Alexa devices bring together the concept of ``interactions with a purpose'' and the concept of ``social mediation'' where two interactions happen simultaneously; one with the device itself and one with the other participants, the latter quite possibly being more important. 

Background input can be an issue as the device's far field microphone often captures input coming from a distance. Essentially, other people in the room who are not participating in the game need to be quiet. Furthermore, there is no clear turn taking mechanism, and participants cannot easily coordinate who should speak at each time. The token-passing workaround from \S\ref{GroupFeedback} was a tentative remedy.
Another possibility would be to use Amazon's ``Echo buttons''.

\section{Summary and conclusions}
We have described a preliminary user study carried out to investigate the Amazon Echo's potential as a CALL platform for children. 
Although the limited scope of the study and the small number of participants mean that conclusions should not be considered as more than suggestive, it seems to us that the core problems we identified are inherent in the basic design of the current Echo and quite serious.

On the positive side, we were interested to see that children often seemed motivated and engaged when other participants interacted at the same time or when they were part of the development loop. If Amazon is able to address the issues we name above, we think Alexa has a great deal of potential as a CALL platform for children.



\section{Bibliographical References}

\bibliographystyle{lrec}
\bibliography{proposal}

\end{document}